\documentclass[12pt,a4paper
]{article} 
\usepackage{graphicx}				
\usepackage{makeidx}
\makeindex 
\usepackage{amssymb}
\usepackage{amsmath}
\usepackage{ulem} 
\usepackage{hyperref} 
\graphicspath{ {figbios/} }
\usepackage {subfigure}
\usepackage{float}
\floatstyle{boxed}
\newfloat{Frame}{h}{boxlist}
\floatname{Frame}{Box}
\newcommand{\aaa}[1]{\noindent\rule{0.03\textwidth}{5mm}\rule[4.5mm]{.94\textwidth}{0.5mm}\rule{0.03\textwidth}{5mm}}
\newcommand{\zzz}[1]{\noindent\rule{0.03\textwidth}{5mm}\rule[0.0mm]{.94\textwidth}{0.5mm}\rule{0.03\textwidth}{5mm}}

\newcommand{\beq}{\begin{equation}}
\newcommand{\eeq}{\end{equation}}

\newcommand{\eeqq}[1]{\lbl{#1} \end{equation} }

\newcommand{\hide}[1]{}

\newcommand{\lbl}[1]{\quad \label{#1} \boxed{ \mbox{#1} }  }

\newcommand{\thistitle}{
 The theory of non-linear oscillator applied to the BiOS mechanism
}
\newcommand{\myhead}{ \fbox{ \thistitle } }
\begin{document}
\sf 
\renewcommand{\familydefault}{cmss}
\renewcommand{\em}{\it}
\large
\title{\thistitle}
\author{
Renzo Mosetti\footnote{OGS, Trieste, Italy, rmosetti@inogs.it} \
 }
\maketitle
\begin{abstract}
The observed pseudo-periodic reversal of the upper layer circulation of the Ionian Sea has been assumed to be related to some internal feedback processes (density driven) by the so called BiOS (Adriatic-Ionian Bimodal Oscillating System) hypothesis. The mechanism seems to be very well described by a non-linear oscillator dynamical system. By setting the state variables as the salinity of Adriatic deep water and the sea level anomaly in the Ionian region a Van der Pol equation has been obtained. The periodic cycle so obtained is of the order of a decade as observed in the data. Furthermore, a periodic term, which mimics the decadal variability of the forcing (mainly atmospheric), produces a period doubling phenomenon which is typical of the forced Van der Pol oscillator.
By considering also a stochastic gaussian noise acting as a forcing and a decadal periodic input term, having an amplitude much smaller than the noise variance, a sub-harmonic stochastic resonance appears in the solution for a certain range of the values of the parameters. This yields the possibility of a longer time scale variability of the BiOS mechanism. 
\end{abstract}
\newpage
\tableofcontents
\newpage
\section{Introduction}
In recent years the hypothesis that the Adriatic-Ionian basin behaves like a bimodal oscillating system between cyclonic and anticyclonic upper layer circulation has been introduced to explain some observational facts.

This mechanism has been called BiOS. (Adriatic-Ionian Bimodal Oscillating System) (Borzelli et al., 2009; Civitarese et al., 2010, Gacic et al., 2010). BiOs is a new different way of explaining the Ionian alternation in the circulation between cyclonic and anti-cyclonic states, because it attributes its causes to internal processes more than to external ones (such as wind  stress).
\begin{figure} [H]
\includegraphics[scale=0.25]{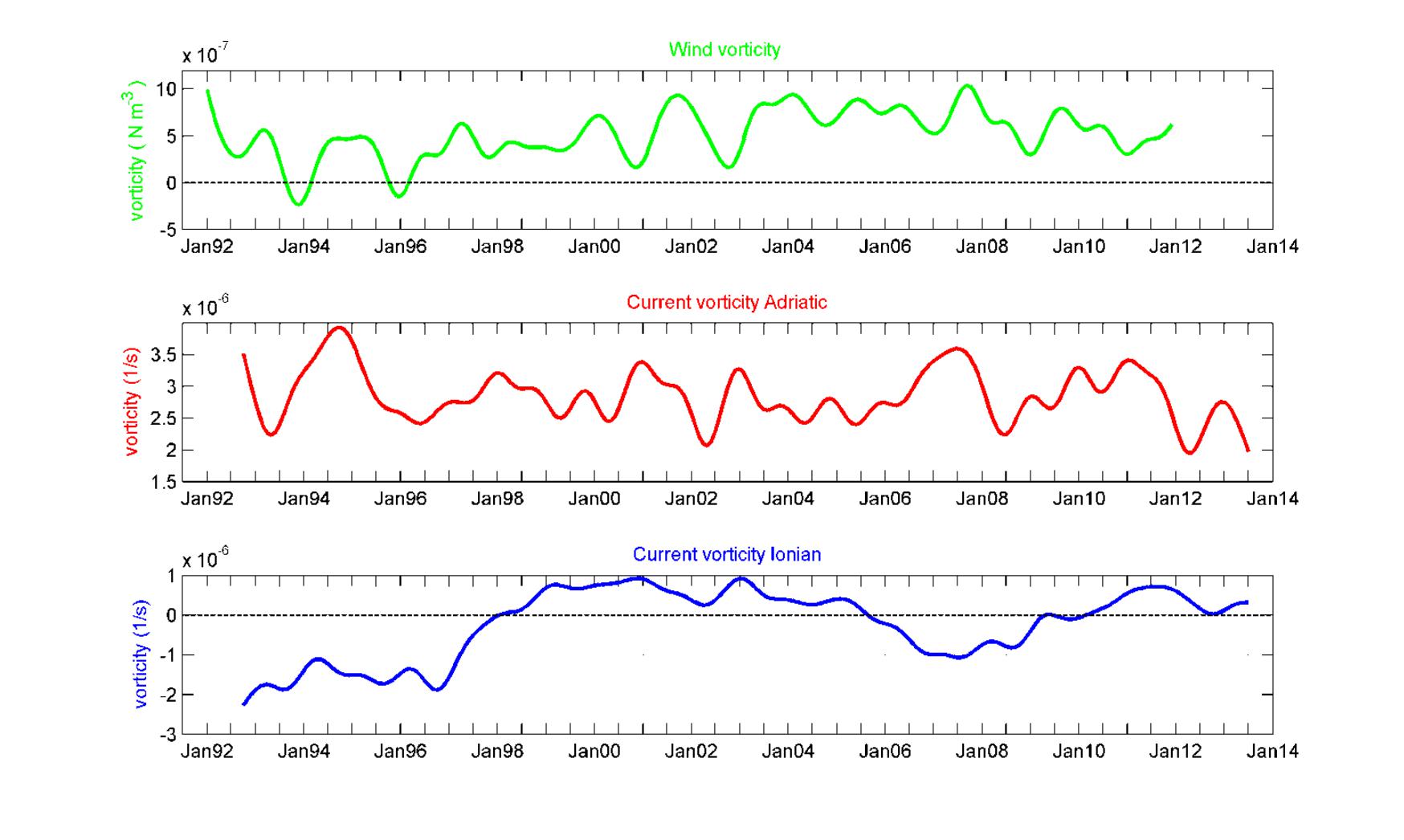}
\label{label}
\caption{Circulation time series in the Adriatic and Ionian basins and wind stress curl. Data have been obtained by satellite altimetry with the geostrophic assumption. }
\end{figure}
\begin{figure} [H]
\centering
\includegraphics[scale=0.6]{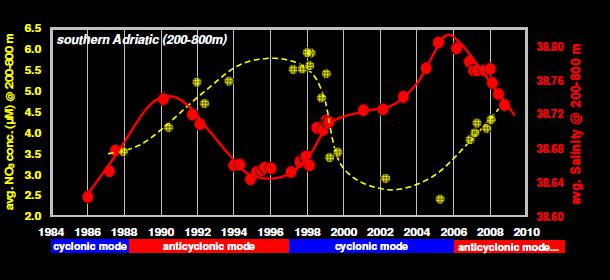}
\label{label}
\caption{The signal of BiOs from Salinity and Nutrient (NO3) data.}
\end{figure}
In fact, looking to Fig.1, it is clear that the wind-stress curl is unable to reverse the circulation. BiOS can give a more general explanation for the phenomenon, based on an “intrinsic” pseudo-periodic oscillation. Note also that the Adriatic general surface circulation is always cyclonic. Also salinity and density data collected in the Southern Adriatic (main source of Eastern Mediterranean deep water) show a decadal variation coherent with the northern Ionian change in the sea level height, and in general, their periodic inversion is compatible with the inversion of circulation. The phenomenology can be described as follow with reference to Fig.3 (left and right). 
\begin{figure} [H]
\centering
\includegraphics[width=.39\textwidth]{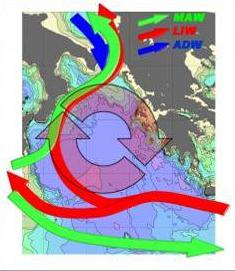}\hfil
\includegraphics[width=.39\textwidth]{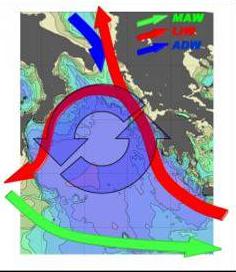}
\caption{The BiOs mechanism (from Civitarese et al., 2010)}\label{3A and 3B}
\end{figure}
The low-salinity MAW (Modified Atlantic Water) advected by the anticyclonic circulation enters the Adriatic basin. This causes a density decrease in Adriatic (since MAW is less salty), so the ADW (Adriatic Deep Water) produced in this period gradually has a lower density than usual. This "light" water spreads towards the Ionian flanks, so that the water level along Ionian flanks gradually increases generating an up-welling of the denser water of the lower layers (and a consequent down-welling of isobars). This leads to a weakening of the anticyclonic upper layer circulation, which at the end gets inverted. Cyclonic regime favors LIW (Levantine intermediate water) to enter the Adriatic, which deep waters are now going to be more salted, thou heavier. As before, denser Adriatic deep water spreads along Ionian flanks increasing their density: this causes a gradual down-welling of the upper lighter density water (and a consequent up-welling of isobars), and so a gradual weakening of the cyclonic regime, that is doomed to be inverted.  Schematically:

Contextualizing to the time interval, in 1994 the anticyclonic gyre extends over the northern Ionian, advecting MAW into the Adriatic (low salinity water): in 1995 anticyclonic gyre starts to become less and less strong and, in 1998, a cyclonic circulation establishes. The salinity in Adriatic basin starts to increase again, since saltier LIW is now advected in. As it is evident in Fig. 3 (left and right). It is important to underline that the anticyclonic trend advects mostly MAW, since the pathway followed by waters is towards the western border of the Ionian, on the other hand, the cyclonic motion induces waters mostly on the eastern part of the Ionian pushing in the Adriatic basin waters saltier than MAW. The inversion happens to be in 1997. Between 1998 and 2005, in fact, a cyclonic gyre is observed, and this favors salty Levantine or Cretan water to enter, inverting the salt content of the Adriatic Deep Water formed until now. Another inversion started in 2006. Actually the circulation seems to be in a cyclonic phase (see again Fig.1).
In a recent paper (Crisciani and Mosetti, 2016) a model has been derived by integrating in the bounded flow domain (the Ionian basin) on the $f$ – plane the two-layer quasi-geostrophic evolution equations. Furthermore, a non linear term which mimics the mutual interaction between the moving layers has been introduced. By considering a stochastic gaussian noise acting as a forcing and a decadal periodic input term, having an amplitude much smaller than the noise variance, a stochastic resonance appears in the solution for a certain range of the values of the parameters. The decadal forcing at the boundary of the Ionian basin can be related to the low frequency atmospheric variability in the surrounding basins. The resonance has a period consistent with the observed decadal time scale of the reversal between cyclonic and anti-cyclonic phases. In this paper, an even more simple dynamical model of a non-linear oscillator is presented. 
\section{The non-linear Feedback-Model}
Essentially the BiOs feedback system could be simplified in terms of a dynamical system with two state variables (Ionian Sea level anomaly $\eta$ and the salinity anomaly $ S $ of the Adriatic Dense Water:\\\
\begin{equation}
\begin{array}{lcl}\dot{S}=aS-b\eta-cS^{3}  \\ \dot{\eta}  =  dS \end{array}
\end{equation}
\\\
According to Fig. 3, this model reflects the following scheme:\\\
    
Entering AMW $\Rightarrow$   Lowering ADW salinity $\Rightarrow$    NO AMW    $\Rightarrow$ Increase ADW salinity$\Rightarrow$  Entering again AMW  \\\ 

So the cycle is starting again.
	The first term on the r.h.s. in the first equation represents the accumulation of the formed Adriatic dense water and it is related the residence time within the Adriatic bottom layer. The second term is the feedback due to the sea level anomaly in the Ionian; the last term is a nonlinear dissipation/mixing process. Note that the cubic exponent arise from the fact that in a power series expansion the cubic term is the first non-linear term changing the sign correctly (if you would use power two the effect would be independent on the fact that an increase of decrease of salinity generates the same negative term). The second equation simply close the feedback which relates sea level anomaly with salinity anomaly.
	Starting with equation 1) we can first of all scaling the time by the typical time scale of the order of months and the sea-level anomaly by the typical depth of the basin. So, we take:\\
$\tilde{t}=t/month;$ \ $\tilde{\eta} =\eta/H$
\begin{equation}
\begin{array}{lcl}\dot{s}=as-b\tilde{\eta}-cs^{3}  \\ \dot{\tilde{\eta}}  =  ds \end{array}
\end{equation}
where 
 After simple manipulations, Equation (2) can be rewritten as:
\begin{equation}
\dfrac{d^{2}s}{d\tilde{t}^{2}}-(a-3cs^{2})\dfrac{ds}{d\tilde{t}}+bds=0
\end{equation}
\\ If the following variable transformation is defined:
$ t = \tilde{t}\sqrt{\dfrac{1}{bd}}$;
$s=pX$;
$\mu=a\sqrt{\dfrac{1}{bd}}$;
$p=\sqrt{\dfrac{a}{3c}}$
\\ The Van der Pol classical equation is obtained:
\begin{equation}
\dfrac{d^{2}X}{d\tilde{t}^{2}}+\mu(X^{2}-1)\dfrac{dX}{d\tilde{t}}+X=0
\end{equation}
\\ The most important characteristics of the VdP equation is that for every $\mu\geq 0$, starting from an arbitrary initial condition, the solution is a stable limit circle. Models of this kind of oscillators have been used for instance, for ENSO dynamics (e.g. J. Sheinbaum, 2002).
 The estimate of the parameters is of course the most critical point in this model. In our case, by using the available data and by considering a residence time for the ADW of the order of 26 months (Vilibic and Supic, 2005), the following estimates have been obtained for the model's parameters in Eq. (2) : (salinity anomaly is a pure number)\\
$a$ residence time of Adriatic Deep Water: 26 months;\\
$c$ estimate from data: $1.13X10^{-9}$ - ($s^{-1}$)\\
$b$ estimate from data: $1.58X10^{-10}$ - ($s^{-1} m^{-1}$)\\
$d$ estimate from data: $2.75X10^{-6}$ - ($s^{-1}$)\\
\\
\begin{figure} [H]
\centering
\includegraphics[scale=0.45]{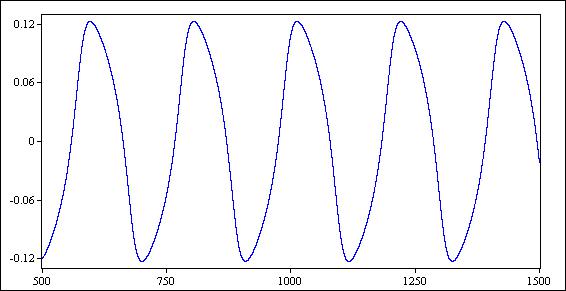}
\label{label}
\caption{Time series of Salinity anomaly (time in months) }
\end{figure}
\begin{figure} [H]
\centering
\includegraphics[scale=0.45]{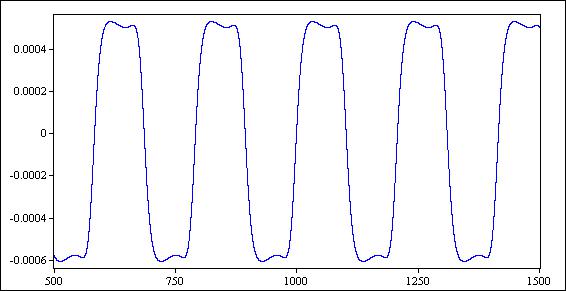}
\label{label}
\caption{Sea level anomaly}
\end{figure}
\begin{figure} [H]
\centering
\includegraphics[scale=0.25]{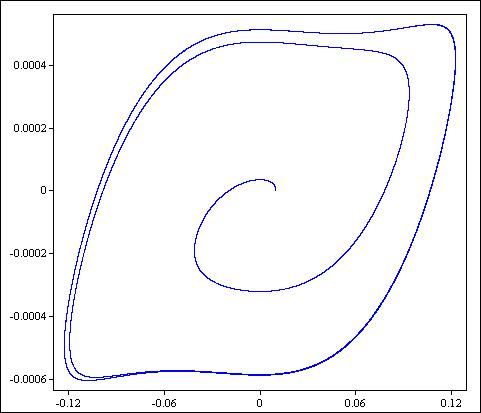}
\label{label}
\caption{Phase space diagram (salinity vs. sea level and the limit cycle}
\end{figure}
Note that the obtained periodicity is of the order of 16 years which is of the same order of the observed reversal of the Ionian circulation. The numerical integration of the equations has been performed by means of a MATLAB code based on a Runge-Kutta ODE high-order solver.
\\
\subsection{The effect of a periodic forcing}
It is well known, since the pioneering works by Van der Pol, 1920  himself and Van der Pol and Van der Mark, 1927, that adding a sinusoidal term to the VdP equation a “noisy” response could arise. More recently it has been proved that the forced VdP equation for certain value of the parameters (mu and omega) exhibit a chaotic behavior through the period doubling cascade process with the loss of periodicity. So, we would like to check if our BiOs model could be also affected by these phenomena within the realistic range of values of the parameters. In the affirmative case this could be interpreted  that the periodicity will be destroyed limiting the validity of the model itself or that, physically, in the presence of an intense periodic forcing the reversal of the Ionian circulation might be stopped.
	The introduction of a decadal periodicity into the system can be supported by the fact that large scale ocean processes are influenced by atmospheric forcing and the appearance of a decadal scale is quite common (Cessi and Louazel, 1992; Wunsch, 2011). Recently,the same process of deep water formation in the Adriatic Sea shows a decadal signal being however related to the atmospheric forcing (Mihanovic H. et al, 2015). 
	The equation by adding a sinusoidal component becomes:\\
\begin{equation}
\dfrac{d^{2}X}{d\tilde{t}^{2}}+\mu(X^{2}-1)\dfrac{dX}{d\tilde{t}}+X=A\cos(\omega t)
\end{equation}
\\Where: $A$ is the amplitude of the periodic forcing and $\omega$ its angular frequancy.
\begin{figure} [H]
\centering
\includegraphics[scale=0.65]{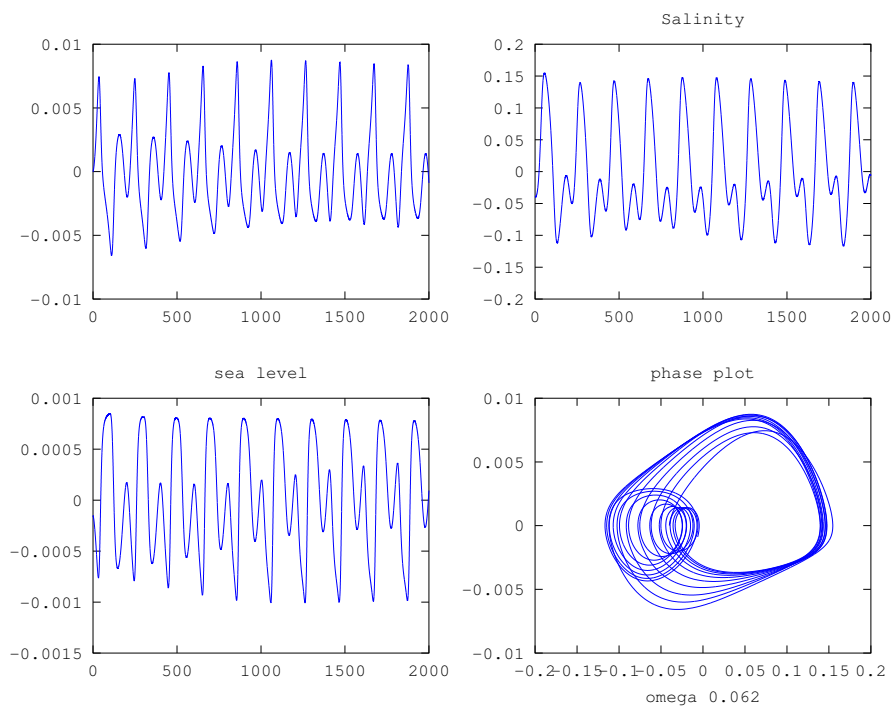}
\label{label}
\caption{Forced Van der Pol oscillator with a decadal periodic function.Upper left: $\dfrac{d \tilde{\eta}}{d\tilde{t}}$}
\end{figure}

\subsubsection{The EMT case}

	We have done a simulation with the hypothesis that during the East Mediterranean Transient (EMT),the effect of an external forcing can be negligible with respect the forcing due to the strong dense water formation within the Aegean basin. So, we stopped the forcing for a period of a decade and than reinserting the periodic forcing after. The results (Fig. 8 and Fig. 9) show a longer period during the EMT and suddenly after a shortening of the periods. If we compare the results with the observed features in Fig. 1 (see the last inversion periodicity from 2006), we get something which seems to be in qualitative agreement. A numerical high resolution hydrodynamic model gives a very similar behavior when only the Aegean deep waters are forcing the Ionian basin (M. Reale, personal communication).
\begin{figure} [H]
\centering
\includegraphics[scale=0.55]{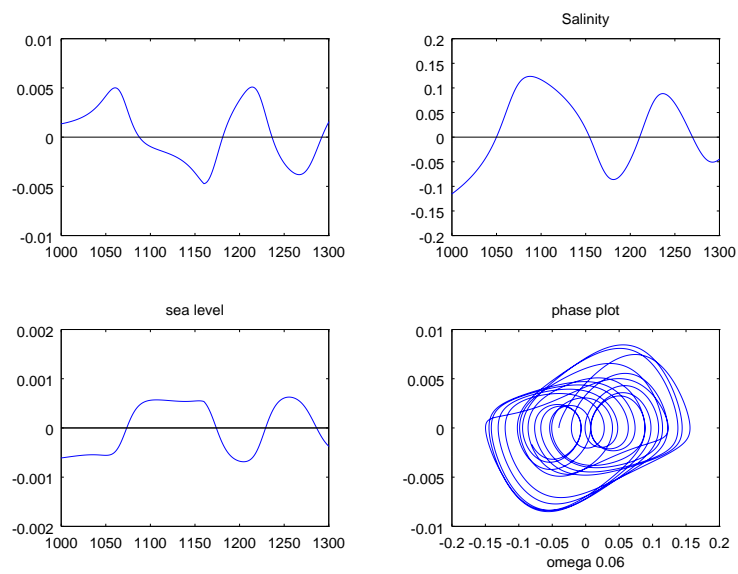}
\label{label}
\caption{Zoom of the solution with a stopped forcing. Panels as in Fig. 7}
\end{figure}
\begin{figure} [H]
\centering
\includegraphics[scale=0.65]{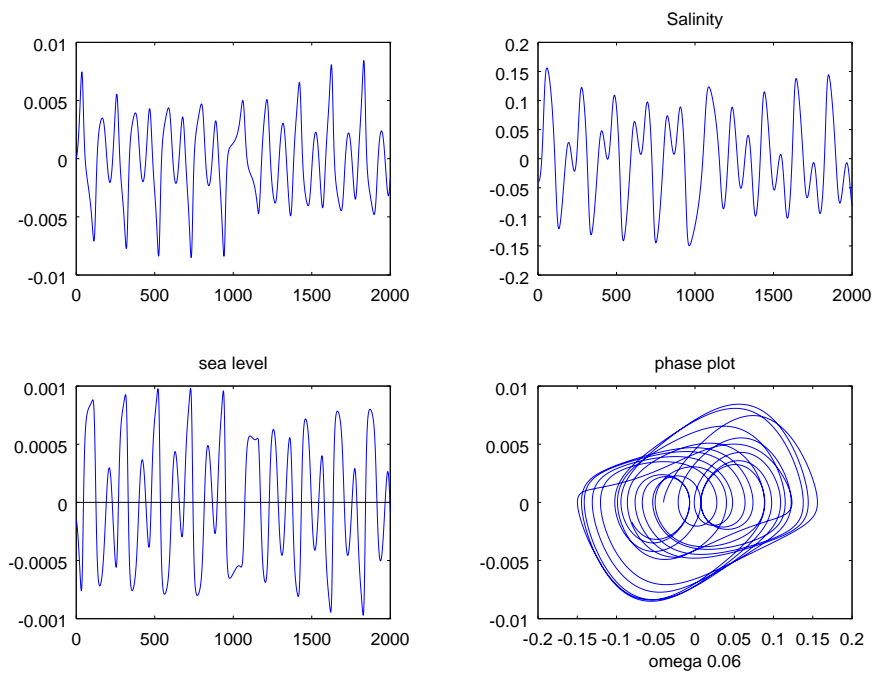}
\label{label}
\caption{Complete solution with a stopped forcing}
\end{figure}
However, the results show that in the limit of the estimated parameter values, the effect of the forcing do not bring the system to a chaotic regime, but possible period doubling could be arise depending on the strength of the forcing. 
\section{The Stochastic Resonance Hypothesis}
The mechanism of stochastic resonance has been introduced in  a seminal paper by Benzi et  al., 1981. Starting from a stochastic differential equation with a periodic forcing, they discovered that even a “small” noise variance (compared to the amplitude of the forcing) may create in the solution a bi-stable response characterized by a rapid transition between the two states with the same period of the forcing term. After the discover, the concept of stochastic resonance has been widely applied to several disciplines in the context of dynamical systems and  in particular to climate dynamics (e.g.  Benzi et al, 1982). 
	The idea is to explore if the dynamical system described by Eq. (5) could exhibit a stochastic resonance when both a periodic decadal forcing and a random noise is introduced. Being the BiOS process supposed to be a bi-stable system, the stochastic resonance could be a candidate model to explain the observed pseudo-periodic reversal of the Ionian vorticity. The noise term can be interpreted as monthly/inter-annual fluctuations of the dense water formation rate. The stochastic resonance has been investigated in the framework of the Stommel thermohaline model (Eyink, 2005). Furthermore, the effect of adding a stochastic term on a box model of the Adriatic-Ionian system has been analyzed recently (Capuano, laurea degree Thesis, 2014). The results show a decadal component in  time evolution of the salinity within the Ionian Sea.
	Following these considerations, we modify the equation as follows:

\begin{equation}
\dfrac{d^{2}X}{d\tilde{t}^{2}}+\mu(X^{2}-1)\dfrac{dX}{d\tilde{t}}+X=A\cos(\omega t)+\varepsilon R(\tilde{t})
\end{equation}
Where: $\varepsilon R(\tilde{t})$ is a Gaussian noise process.

A large set of runs have been done by small changes in the variance of the random noise since the stochastic resonance usually appears within a very narrow range of variance. 
\begin{figure} [H]
\centering
\includegraphics[scale=0.65]{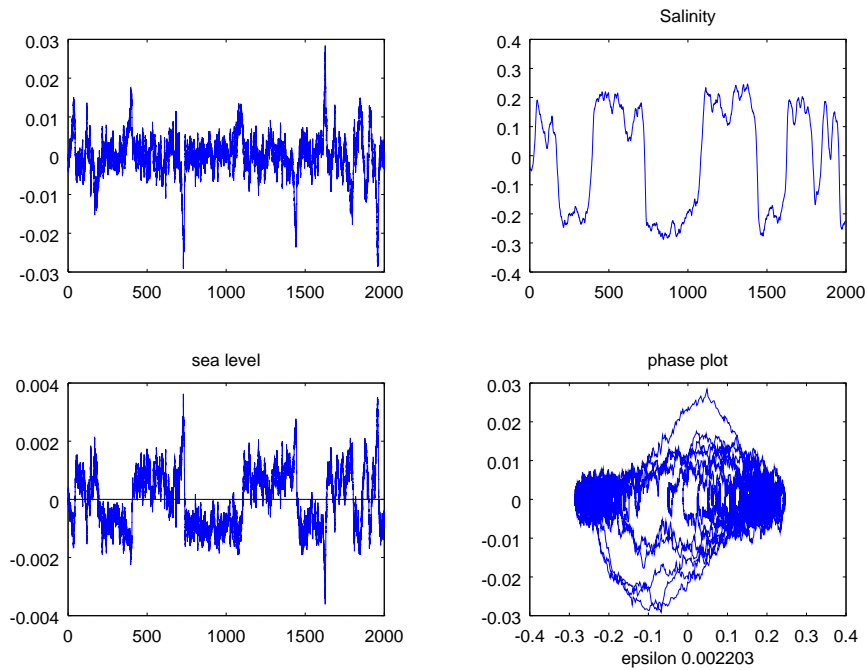}
\label{label}
\caption{Sub-harmonic stochastic resonance. Panels as in the previous figures. Note the two equilibria in the phase plane}
\end{figure}
It is very interesting that we have obtained what is called a sub-harmonic stochastic resonance rather than a resonance with the same period of the forcing. This has been found in certain dynamical systems (Chialvo et al., 2002).	
	This fact leads to the possibility of the existence of a longer time scale component in the circulation inversion in the Ionian depending of the inter-annual variability of the forcing that could be related to the inter-annual variability of the dense water formation in the Adriatic Sea.

\section{Discussion and conclusions}
The driving mechanisms behind the decadal reversal of the Ionian Sea upper layer circulation raised a considerable debate in the Mediterranean scientific community. It is still unclear what is the driving force. It has been suggested that the reversal can be driven by variations in wind stress curl over the basin, baroclinic dynamics acting within the Adriatic-Ionian System (AIS) or baroclinic dynamics driven by thermohaline properties at the AIS eastern boundary. The variability observed in the Ionian circulation has been often associated with a change in the wind stress curl over the area ( Pinardi et al.,1997, Pinardi et al., 2015). Recently, Theocharis et al. 2014 suggested that the decadal variability observed in the Ionian circulation reflects an internal mechanism driving the alternation of the Adriatic Sea and the Aegean Sea as main DWF (Deep Water Formation) sites for the Eastern Mediterranean (EMED). This thermohaline pumping (Theocharis et al., 2014 ; Velaoras et al.,2014) involves the whole thermohaline cell of EMED and not only the Adriatic-Ionian Sea as in the BiOS hypothesis (Gacic et al.,2010). In both cases the role of long term atmospheric forcing has been consider as negligible. In a recent paper by Mihanovic´ et al, 2015 it is  claimed that BiOS is the dominant generator of the Adriatic decadal variability. 
Despite its simplicity, this dynamical model gives results that are in agreement with the observed reversal of the Ionian circulation. It is in the assumption of the BiOS phenomenology an intrinsic role of a feedback process and this is captured by the oscillator dynamics. The results however, seems to re-introduce the role of atmospheric forcing (and the related effects on deep water formation) as an fundamental factor for the reversal of the Ionian circulation: even the Van der Pol oscillator needs an active source of energy for alimenting the negative resistance (a triode) necessary to sustain the periodic output.
\paragraph{Acknowledgment}
The author is indebted to dr. Fulvio Crisciani for useful discussion on this subject.\\
\section{References}

Benzi R., A. Sutera and A. Vulpiani, 1981: The Mechanism of stochastic resonance. J. Phys. A: Math. Gen., 14 L453-L457.

Benzi R., G. Parisi, , A. Sutera and A. Vulpiani, 1982: Stochastic resonance in climate change. Tellus, 34, 10-16.

Borzelli, G. L. E., Gacic M., Cardin, V., and Civitarese, G., 2009: Eastern Mediterranean
transient and reversal of the Ionian Sea circulation, Geophys. Res. Lett.,
36, L15108, doi:10.1029/2009GL039261.

Capuano C., 2015: Non-linear box models for the study of the interannual variabuility of the thermohaline parameters of Adriatic-Ionian basin. University of Trieste, Inter athenaeum Master’s Degree in Physics - Earth and Environmental Physics Curriculum,  Academic Year 2013-2014. Supervisor: Renzo Mosetti.

Cessi, P., Louazel, S.,1992: Decadal Oceanic Response to Stochastic Wind Forcing,
Journal of Physical Oceanography, Vol.31, 3020-3029.

Dante R. Chialvo, Oscar Calvo, Diego L. Gonzalez, Oreste Piro,and Guillermo V. Savino, 2002:
Subharmonic stochastic synchronization and resonance in neuronal systems. PHYSICAL REVIEW E, VOLUME 65, 050902

Civitarese, G., M. Gacic, G. L. Borzelli and M. Lipizer, 2010: On the impact of the Bimodal Oscillating System (BiOS) on the biogeochemistry and biology of the Adriatic and Ionian Seas (Eastern Mediterranean). Biogeosciences, 7, 3987-3997.

F. Crisciani and R. Mosetti, 2016: Is  the bimodal oscillating Adriatic-Ionian circulation a stochastic resonance ?. BGTA,Vol. 57, 275-285.

Eyink G. L., 2005: Statistical hydrodynamics of the thermohaline circulation in a two-dimensional model. Tellus, 57A, 100-115.

Gacic, M., G.L.E. Borzelli, G. Civitarese, V. Cardin and S. Yari, 2010. Can internal processes sustain reversals of the ocean upper circulation? The Ionian Sea Example. Geophys.Res. Lett., 37, L09606.

Mihanovic H. et al, 2015: Mapping of decadal middle Adriatic oceanographic variability and its relation to the BiOS regime. JGR-Oceans DOI: 10.1002/2015JC010725.

Pinardi, N., Korres, G., Lascaratos, A., Roussenov, V., Stanev, E., 1997: Numerical simulation of the interannual variability of the Mediterranean Sea upper ocean circulation. Geophys. Res. Lett. 24, 425-428.

Pinardi N.,  M. Zavatarelli, M. Adani , G. Coppini, Claudia Fratianni ,P. Oddo, S. Simoncelli , M. Tonani, V. Lyubartsev , S. Dobricic and A. Bonaduce, 2015: Mediterranean Sea large-scale low-frequency ocean variability and water mass formation rates from 1987 to 2007: A retrospective analysis. Progress in Oceanography 132, 318–332

J. Sheinbaum, 2002: Current theories on El Niño-Southern Oscillation: A review. Geofísica Internacional (2003), Vol. 42, Num. 3, pp. 291-305.

Theocharis, A., G. Krokos, D. Velaoras and G. Korres, 2014: An internal mechanism driving the alternation of the Eastern Mediterranean dense/deep water sources, In The Mediterranean Sea: Temporal Variability and Spatial Patterns, edited by G. L. E. Borzelli, et al., AGU Geophys. Monogr. Ser., 202, pp. 113–137, John Wiley, Oxford, U. K., doi:10.1002/9781118847572.ch8., 

Velaoras, D., Krokos G., Nittis K., and Theocharis A., 2014: Dense intermediate water outflow from the Cretan Sea: A salinity driven, recurrent phenomenon, connected to thermohaline circulation changes, J. Geophys. Res. Oceans, 119, 4797–4820, doi:10.1002/2014JC009937,2014

B. van der Pol, 1920: A theory of the amplitude of free and forced triode vibrations,
Radio Review, 1, 701-710, 754-762.

B. van der Pol and J. van der Mark, 1927: Frequency demultiplication, Nature, 120, 363-364.

I. Vilibic and N. Supic, 2005: Dense water generation on a shelf: the case of the
Adriatic Sea, Ocean Dynamics,55, pp. 403–415.

Wunsch, C., 2011: The decadal mean ocean circulation and Sverdrup balance. J. Mar. Res., 69, 417-434.

\
\hide{ 
} 
\end{document}